\documentclass[UTF8,a4paper]{article}
\usepackage{amsmath}
\usepackage{graphicx}
\usepackage{subfigure}
\usepackage{epstopdf}
\usepackage{geometry}
\begin{document}

\title{\bf Analysis of the behavior of singlet pairs in inorganic crystal XOCl(X=Ti,Bi)}
\author{Chen-Huan Wu
\thanks{chenhuanwu1@gmail.com}
\\College of Physics and Electronic Engineering, Northwest Normal University, Lanzhou 730070, China}

\maketitle
\vspace{-30pt}

\begin{abstract}
\begin{large}
This article explore the behavior of singlet pairs in Inorganic crystals and take the Ti-Ti and Bi-Bi dimer for example, including the transition in critical temperture by directly or indirectly.
Through the analysis, It is proposed that with the decrease of temperature, the strength of spin-orbit coupling (SOC) increases and the phase difference also increases.
In the one-dimensional spin 1/2 chain system of TiOCl and BiOCl crystals, the possibility of reversible parameter modulation is proposed by 
the calculation of the first principle analyzing the process and analysis of the structural phase transition process.
It's shown that the different sturcture transition is relate to the difference of the dimer structure and the fluctuation of orbital order 
through the comparative study of TiOCl and BiOCl.

\end {large}
\end{abstract}

\begin{large}
\section{INTRODUCTION}
The critical temperature of phase transition of the inorganic crystal TiOCl has been observed as $T_{c1}=67 k$ and $T_{c2}=90 k$ , and the two transitions
are one order and two order transition, respectively \cite{Shaz M}. 
The former is accomplished by the two-fold superstructure of Ti-Ti singlet pairs in a layer-shape SU(N) square
lattice, and since it's the incommensurate spin-Peierls state or the commensurate dimerized state for $N\ge 5$ \cite{Harada K}, there is dominated by the spin-Peierls
state in low temperature through the first order transition in $T_{C1} $and resulting in the existence of spin gap in intermedia phase, 
the gap is becomes the pseudospin with the increase of temperature. 
The first order transition is accompany with a huge change of magnetic behavior which produce a magnetic gap below $T_{c1}$ and dominated by a low symmetry 
monoclinic $P2_{1}/m$ phase \cite{Blanco-Canosa S}. The transition to the commensurate dimerized state is happen within the $T_{c1}$ and $T_{c2}$.
These behavior is indeed the Ti-Ti dimerization modulated by the temperature.
As for the semiconductor BiOCl which has the similar propertice in chemical and structure with TiOCl but with more higher symmetry phase $P4/nmm$, 
we can also see the important role of the Bi-Bi dimer in the structural transition.
The singlet pairs of Ti-Ti dimer and Bi-Bi dimer is horizontal presented in the Fig.1, they are the view of ab plane with the top-to-down layer-struture, 
by increase the degrees of spin-orbit coupling (SOC), the amplitude of bond \cite{Wu C} is enlarged and therefore the collective dynamics is more severe \cite{Hu F Q}.
Under this condition, the in-phase array singlet pairs is possible to transition to the out-of-phase arrangment, this is the resonce why it's a good platform to 
study the propertices of solid material. I also indicate and compare the dimers behavior of TiOCl and BiOCl to see whether the Bi-Bi dimers in BiOCl have the same 
propertices as TiOCl. The reason why using layer-shape TiOCl and square-shape BiOCl is these two crystals have similar electron configuration (in ab plane), 
and they show a similar nature in dimers, magnetic frustraction, and the fluctuation effects \cite{Ma F}. They have the similar peak-position of the Raman spectra in room 
temperature\cite{Lemmens P,Biswas A} as well as the similar band gap (see below).

\section{MODEL AND ANALYSIS}

Unlike the VBS state which is temperature-independent if non-doping \cite{Wu C}, the spin-Peierls transition which relate to the obvious changes
is found in TiOCl at low temperature ( 65 k $\sim$ 67 k, \cite{Shaz M},
26 k$\sim$ 100 k, \cite{Seidel A} ) even if subtract the Curie tail.
Fig.1(a) show the layered struture of TiOCl, we can see three layers stack up and down along the a axis, and the distance between each layer is 1.8915\AA\ acording
to data of Ref.\cite{Shaz M}. 
Fig.1(b) show the irregular octahedron formed in TiOCl crystal, and (c) is the irregular octahedron decahedron formed in BiOCl crystal \cite{Zhao J}.
According to the data of Ref.\cite{Seidel A},
the nearest Ti-Ti ions pair is the one belong to different layer which mearsured as 3.21 \AA\ \cite{Shaz M} (i.e., the Ti ions in site 1 and site 2 which showed in Fig.1(a)).

In the model shown in the Fig.2, there are two kinds of interaction, one of it is the
superexchange interaction (orbit $d_{xz}$) along the a axis which via the oxygen orbitals and another is the directly exchange interaction from the chains in layer plana and formed by Ti-Ti
singlet pairs along the b axis (orbit $d_{xy}$) which take a key role in the process of crystal sturcture transition.
Both these two exchange interaction 
together to complete the phase transition process, i.e., containing both the interchain and intrachain interaction.
Since the crystal field (see Fig.1(a)) split the d level into the excited one $d_{xy}$, $d_{yz}$, and $d_{xz}$ \cite{Seidel A}, 
it's conscious for what exhibited in Fig.3. 
Despite this, the variation of distence within Ti-Ti (along b axis) is nearly three times large than that of the nearest Ti-Ti\cite{Shaz M}.
That also reflect that there is a great difference in the strength of the exchange between these two interaction.

The ab plane of TiOCl crystals which showed in Fig.2 is consist of the Ti-Ti dimers, the Fig.2 reflect the existence of tight-binding
chains in different layers with almost the same distance.
Fig.2(b) shows the ab plane of  BiOCl, it's obviously that the crystal of BiOCl is present as a square-shape due to it's tetragonal phase,
and the distence of Bi-Bi spin pair which consisted of the majority and minority spin is not much difference.
It's shown that  BiOCl is more stable that TiOCl due to its' superlattice order suggested by the XRD result\cite{Zhou W}.
Spacing of Ti-Ti dimer is shorter than that of 
Bi-Bi dimer according to the crystal structure models I using here (The spacing of Ti-Ti is 3.4653\AA\ and for Bi-Bi is 3.9032\AA\ here which are close to 3.3415\AA\
and 3.980\AA\ from Ref.\cite{Shaz M} and Ref.\cite{Zhang K L} respectively). The distance within 
each singlet pair and that between two pairs is not equal, but since the difference is within 0.1\AA$\sim 0.2$\AA in most case, so it's ignorable.
So Hamiltonian of the Heisenberg ladder model which shown in the Fig.1 can be written as  

\begin{equation}   
\begin{aligned}
H=-J_{1}\sum_{\langle i,j \rangle}{\bf S_{i}\cdot S_{j}}-J_{2}\sum_{\langle i,j \rangle \langle k,l \rangle}({\bf S_{i}\cdot S_{j}})({\bf S_{k}\cdot S_{l}})
 -J_{3}\sum_{\langle i,j \rangle \langle k,l \rangle \langle m,n \rangle}({\bf S_{i}\cdot S_{j}})({\bf S_{k}\cdot S_{l}})({\bf S_{m}\cdot S_{n}})
\end{aligned}
\end{equation}

Where $J_{1}$, $J_{2}$, and $J_{3}$ is the coupling constant in this two-, four-, and six-spin configuration and ${\bf S}$ is the spin operator of different
site, and the minus sign represents antiferromagnetic. 
In fact it's the zigzag chains if takes the view of ac plane of TiOCl.
The spin operators in $(\cdot\cdot\cdot)$ describe the correlation of singlet pair. For this ${\bf S}=1/2$ system $H_{ik}$ the singlet projection is 
$\langle {\bf S_{i}\cdot S_{j}} \rangle=\frac{1}{4}-{\bf S_{i}\cdot S_{j}}$ \cite{Beach K S D}, and the strength of interaction between singlet pairs $\langle i,j \rangle$ is
$U_{ij}=2C\int|\phi_{i}(x)|^{2}|\phi_{j}(x)|^{2}d^{3}x$ where $C$ is interaction constant and $\phi$ is the wave function of these two sites \cite{Wu C}.
It's the similar form for orther singlet pairs.
This ladder model(Equ.(1)) conclude the effect of itinerant electrons which is not localized and have contribute to the spin-Peierls transition,
while the localized electron occupy the degenerate orbit states.
The ${\bf S_{i}}$ in Equ.(1) has ${\bf S_{i}}=\frac{1}{2}\sum c_{i\sigma}^{\dag}\mathcal{\rho^{\mu}}c_{i\sigma}$ where $\sigma$ denotes the spin and 
$\mathcal{\rho^{\mu}}$ is the pauli matrix with $\mu=x,y,z$. $c_{i\sigma}^{\dag}$ and $c_{i\sigma}$ are the creation operator and annihilate operator in site i
respectively which has $c_{q\sigma}=\frac{1}{\sqrt{N}}\sum_{i}e^{iqr}c_{i\sigma}$ and $c_{q\sigma}^{\dag}=\frac{1}{\sqrt{N}}\sum_{i}e^{-iqr}c_{i\sigma}^{\dag}$\cite{Hubbard J}
where $q$ stands the momentum space of Brillouin zone,

The repretation of this three singlet pairs system in TiOCl using the hopping parameter is

\begin{equation}   
\begin{aligned}
H=-t_{1}(c^{\dag}_{i\sigma}c_{j\sigma}+c^{\dag}_{j\sigma}c_{i\sigma})-t_{2}(c^{\dag}_{k\sigma}c_{l\sigma}+c^{\dag}_{l\sigma}c_{k\sigma})
 -t_{3}(c^{\dag}_{m\sigma}c_{n\sigma}+c^{\dag}_{n\sigma}c_{m\sigma})+U\sum_{r=i\sim n} n_{r\sigma_{\uparrow}}n_{r\sigma_{\downarrow}}
\end{aligned}
\end{equation}

where the last term is the site antiferromagnetic interaction (Coulomb repulsion), and this term equal to zero for the zero modes.
Here ignore the weak interaction of ferromagnetic between different layers and the third particle in same layer.
according to Ref,\cite{Lemmens P} the hopping $t$ of TiOCl for nearest neighbor atoms (Fig.2(a)) in one layer is -0.21 which is about seven times of the hopping 
between first atom and third atoms and also the same multiple compare to hopping along a axis.
Further, since the special tructure of BiOCl which exhibit a square shape in the view of ab plane and a hexagon shape in the bc plane, 
the caculation of square lattice model and honeycomb lattice model has been presented in the Ref.\cite{Wu C} and Ref.\cite{Baskaran G}, respectively.

The ab plane shown in Fig.2(a) only exhibit the directli interaction intrachains, the interchains one is ont shown. 
Since the coupling strength $J$ (discrebe the exchange interaction) fot TiOCl of intrachain is much large than that of interchain's which is about 300 times large
than the latter one \cite{Blanco-Canosa S}, and for BiOCl this multiple is bigger (according to the DOS pattern). The coupling $J=2t^{2}/U$ for antiferromagnetic
exchange\cite{Blanco-Canosa S} where $t$ is the hopping between nearest spin antiferromagnetic pairs. According to this we can know that the on site interaction $U$
for Ti-Ti dimer which along the b axis is an order of magnitude smaller than the Ti-Ti in in adjacent layers.

It's the two-fold superstructure which appear below
$T_{c1}$ for TiOCl. Since the superstructure parameter can be separated into basic structure parameter \cite{Palatinus L}, and the modulation parameters can be applied in adjustion
in amplitude and interaction strength of each atom in this dimers system. This two fold structure can lead to two independent atoms in supercell \cite{Palatinus L},
i.e., the spin singlet pair, and it's found that the interaction strength of the pairs atoms is doubled in the first transition point for TiOCl \cite{Shaz M} 
and indicate that the two phases which dominate before and after the phase transition are direct coexistence, that's consistent with the characteristics of 
first order transition \cite{Banerjee A}.

\section{Simulation Results and Discussion}

The simulation and the first-principle calculations are performed use the method of  Vienna ab initio simulation package (VASP).
The split of crystal field U-term in LDA$+$U caculation (LDA, local-density-approximate) is relate to the spin arrangement and the occupation of bands, 
to caculate the antiferromagnetic interaction between the singlet pairs, the U-term can be set in the orbitals which occupying by the interacting atoms (i.e., the
d orbital here), and that will lead to split of $d_{xy}$ band along the b axis (not shown here).
That's clearly reflected in the transition of 
in-phase pattern and out-of-phase pattern (Fig.2) as well as the peak distributi on density of states (DOS) graph (Fig.3). By analyzing Fig.3, it's obvious
that the density of total state is simplely the summation of the each orbital state's density, and the peak of Bi-d's state density is much large than that of Ti-d's.
That means the interaction strength within the pairing atoms of Bi-Bi is much large than that of Ti-Ti. 
In Fig.3(c) and (d), we can see that the peaks of O-p and Cl-p is in the left side and far away from the peak of Ti-d, and for BiOCl the peaks of O-p and Cl-p is in the 
right side and far away from the peak of Bi-d (This phenomenon is also reflected in the bound struture of Fig.4). 
And the band gap width of BiOCl is also much large than the TiOCl's due to semiconductor characteristics of BiOCl.
To indicate the different position of these two mainly peak in Fig.3(c) and (d), I present the DOS of $Bi_{1-x}Ti_{x}OCl$ with $Bi:Ti=1:1$.
In this graph, the peaks of all lines have fallen compare to above ones, and the peaks of O-p,Cl-p, and Ti-d are slightly moving to the Fermi level.
This exhibit the reduction of the energy gap and the interaction between O-p,Cl-p, and Ti-d is increase. 

In Fig.4, the LDA band structure of BiOCl is shown, the two subgraphs in the left side with band gap of 2.186 eV is the case of without spin-orbit coupling (SOC),
and the two subgraphs in the right side with band gap of 2.107 eV is the case with SOC. The band sturcture has already presented in the Ref.\cite{Seidel A} and I don't 
repeat here. It's easy to find that the band sturcture of both the TiOCl and BiOCl have no band touching point, and the band gaps of these two crystal is almost equal.
For a stable coupling, the weak SOC corresponding to Gaussian distribution of atoms and the strong one corresponding to non-Gaussian distribution \cite{Hu F Q},
and the simulation results from Ref.\cite{Hu F Q} also show that Gaussian distribution is broken with the increase of strength of SOC, and the system become 
out-of-phase, and the spin polarization ratio of these two spin in a singlet pair $\mu_{s}=\frac{n_{\uparrow}-n_{\downarrow}}{n_{tot}}$ 
is also increse if the strength of SOC is large enough, where $n_{\uparrow}$ and $n_{\downarrow}$ is the majority spin atom (here is the up-spin) and the minority
spin (here is the down-spin) atoms' number and $n_{tot}$ is the total number of atoms, and a violent high-frequency vibration signal appeared\cite{Hu F Q}.

Since the appearence of antiferromagnetic spin interaction is due to the change of electron structure essentially, the disturbution of density of electron also reflect
the crystal sturcture and play a key role in the spin, charge, and magnetic order as well as the phase transition, and that's also how crystal field
Accroding to analysis of TiOCl's transition, I expect that the strength of SOC of TiOCl is increase as the temperature decrease, and therefore the order properties
is also weakened as the temperature decrease.
That means in low temperature the spin polarization ratio is more large and therefore the spin long range order is hard to establish, a evident is the 
credible signal of nuclear magnetic resonance (NMR) under low temperature \cite{Li Y Q}. 
The orbital fluctuation rise with the temperature rise, which consistent with the previous inferences that too much orbital fluctuation 
is not conducive to the increasement of SOC, will gives rise to entropy, and enhance the spin ferromagnetic interaction between adjacent layers along the a direction
(see Fig.1) \cite{Khaliullin G}.
The fluctuation also gives rise to a pseudo spin gap above $T_{c2}$ which no shown in the Fig.3\cite{Shaz M}.
The broad NMR signal also reflect the incommensurate propertice of spin-peierls state, and for TiOCl, according to Ref.\cite{Krimmel A,Shaz M}, a spin gap of 
430K which is observed from NMR\cite{Krimmel A} is exist as a intermediate phase separated from the undimerized paramagnetic state by the critical temperature $T_{c2}$ \cite{Krimmel A},

To explore the difference of these two dimers in Fig.2, the X-ray drffraction (XRD) pattern of doped crystals $Bi_{1-x}Ti_{x}OCl$ is presented in the Fig.5, 
the different lines corresponding to different doping ratio (here using the wt.$\%$). Note that evry lines in the XRD pattern was raised by ten units for
more convenient to theoretical description.
Roughly, with the increase in Ti content, the peak value of the XRD is also increased, e.g., the highest peak of ratio $48.20\ wt.\%$ is near one hundred,
and we can see that the peaks are mainly concentrated in two regions of 10 deg. and 30 deg..

In conclusion, since for both the TiOCl and BiOCl the singlet pairs in incommensurate structure phase can be modulated by the single modulation \cite{Elcoro L}
(include both the temperature modulate and pressure modulate\cite{Zhao J}), 
In principle, the structure distort effect of SOC is through the way of enhance the phase difference and change the electron configuration.
It provide a good platform to indicate the SOC effect by the behaviours of singlet pairs with spin antiferromagnetic consist if the up-spin and down-spin
(or the majority and minority spin). This effect which is obviously in the TiOCl crystal structure provide a new way of structure transition modulation by
the phase difference modulation through the adjustion of harmonic potential in the laboratory.

\begin{figure}[!ht]
   \centering
   \begin{center}
     \includegraphics*[width=0.8\linewidth]{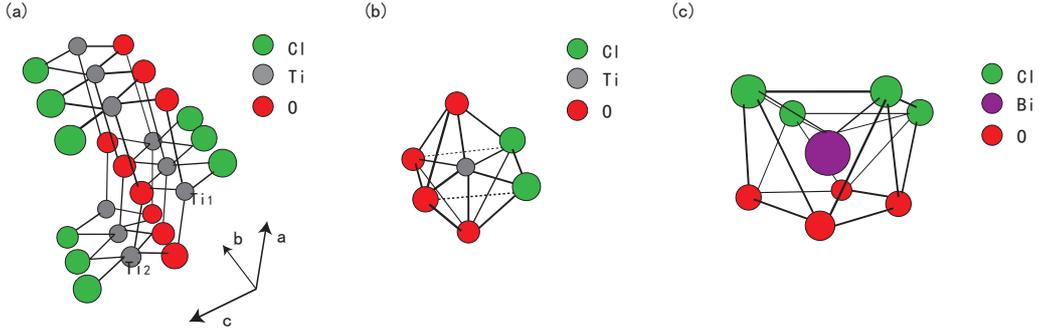}
   \caption{(a) is the schematic views of crystal struture of TiOCl. The indicator in lower right corner is the three axes indicate the a-, b-, c-direction 
which perpendicular to layer, parallel to the Ti ions chain in layer, and perpendicular to the Ti ions chain in layer, respectively.
 (b) and (c) are the irregular polyhedron formed in TiOCl and BiOCl crystal, respectively.}
   \end{center}
\end{figure}

\begin{figure}[!ht]
   \centering
     \subfigure[]{\includegraphics[width=0.8\linewidth]{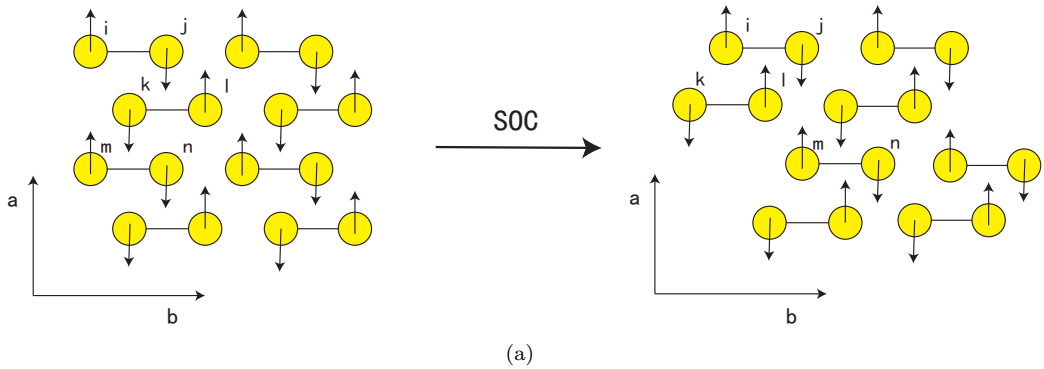}}
     \subfigure[]{\includegraphics[width=0.4\linewidth]{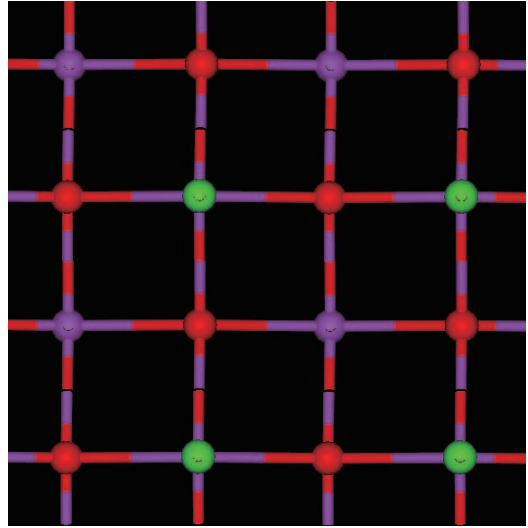}}
   \caption{(a) is the singlet pairs in ab plane of TiOCl crystals. The left graph is the Ti-Ti dimers under in-phase situation,
and the right graph is that under out-of-phase situation. The arrow in the middle reflect this transition is effected by the degrees of spin-orbit coupling (SOC).
(b) is the ab plane of BiOCl.}
\end{figure}

\begin{figure}[!ht]
   \centering
   \subfigure[]{\includegraphics[width=0.4\linewidth]{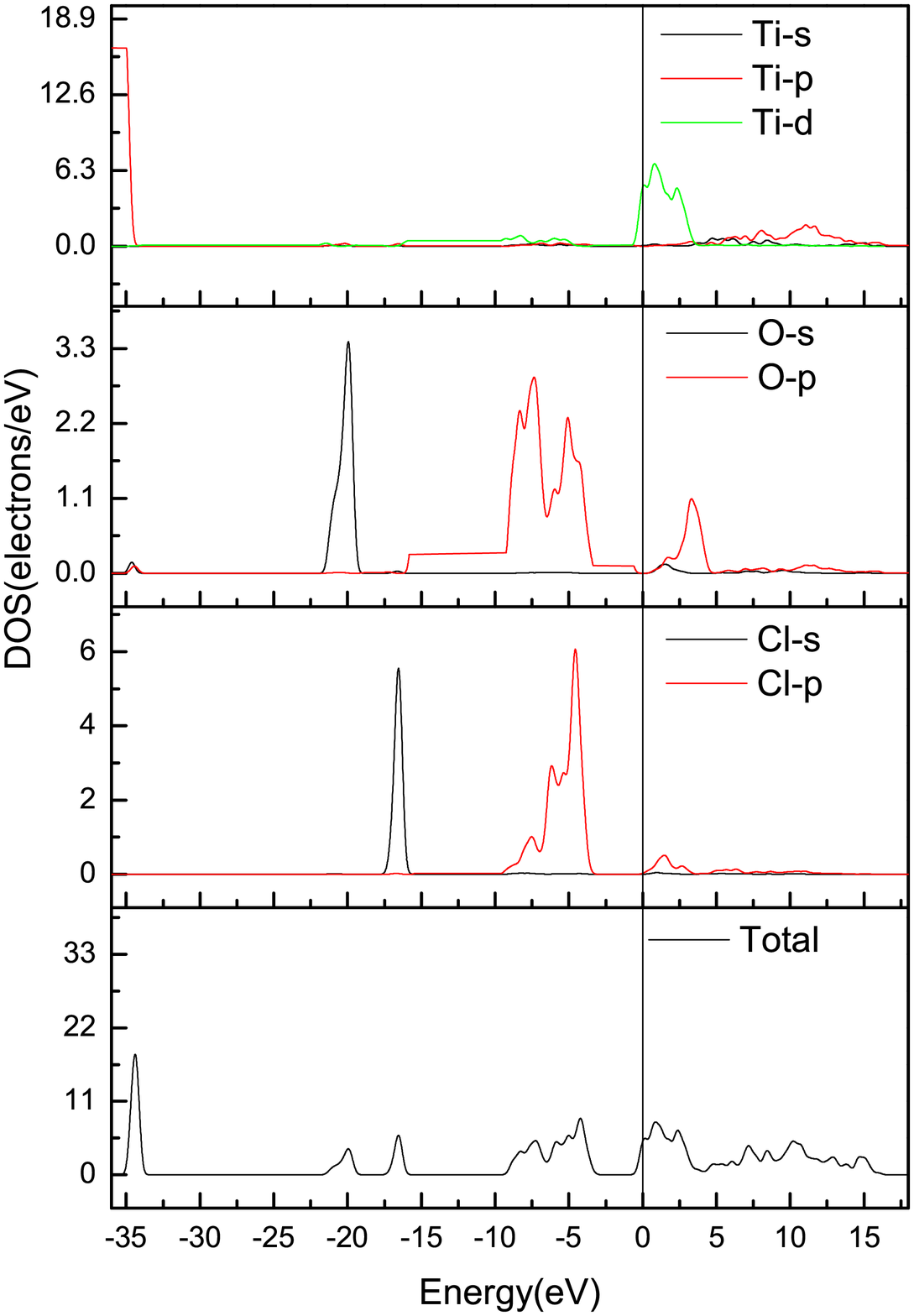}}
   \subfigure[]{\includegraphics[width=0.4\linewidth]{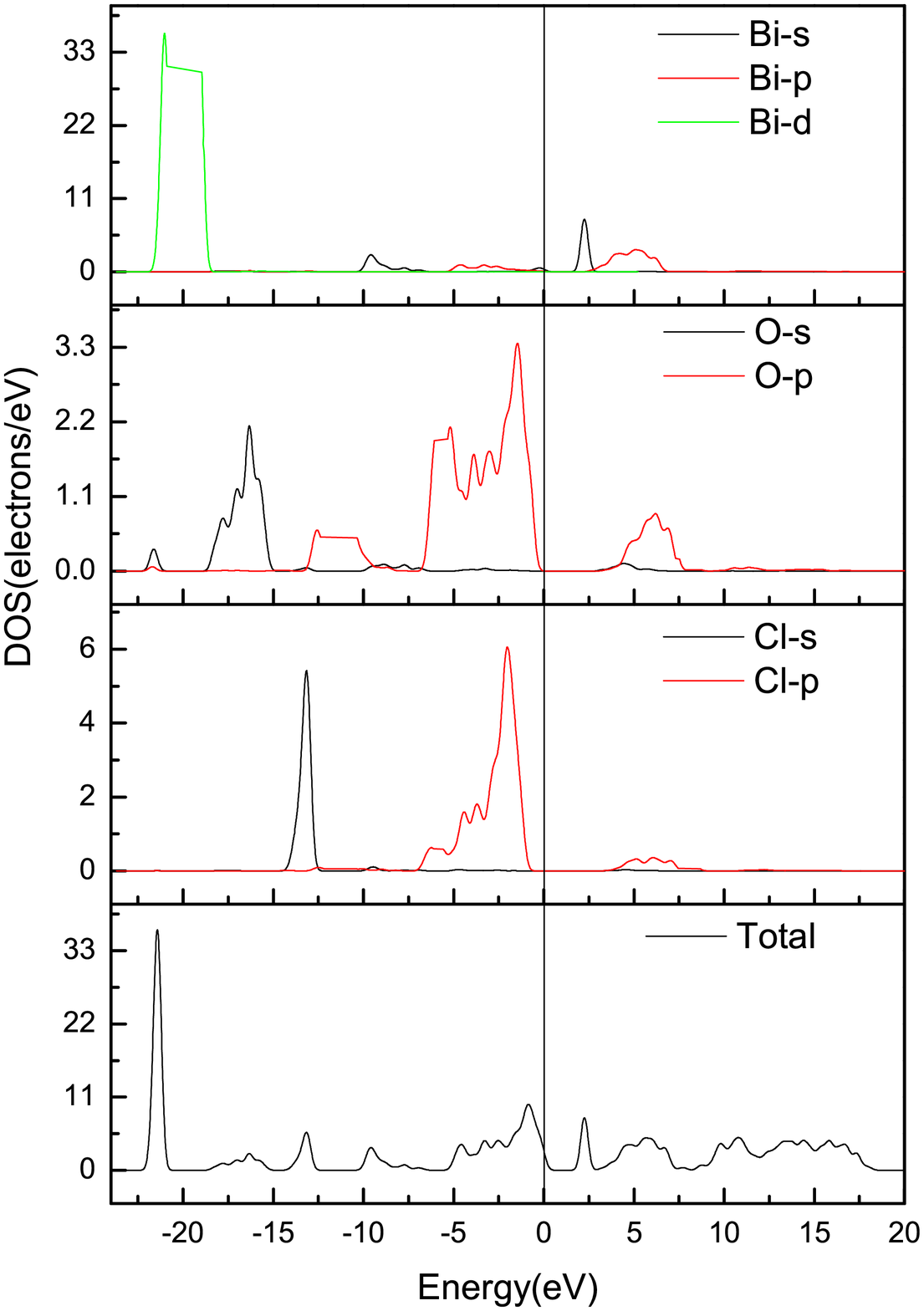}}
   \subfigure[]{\includegraphics[width=0.4\linewidth]{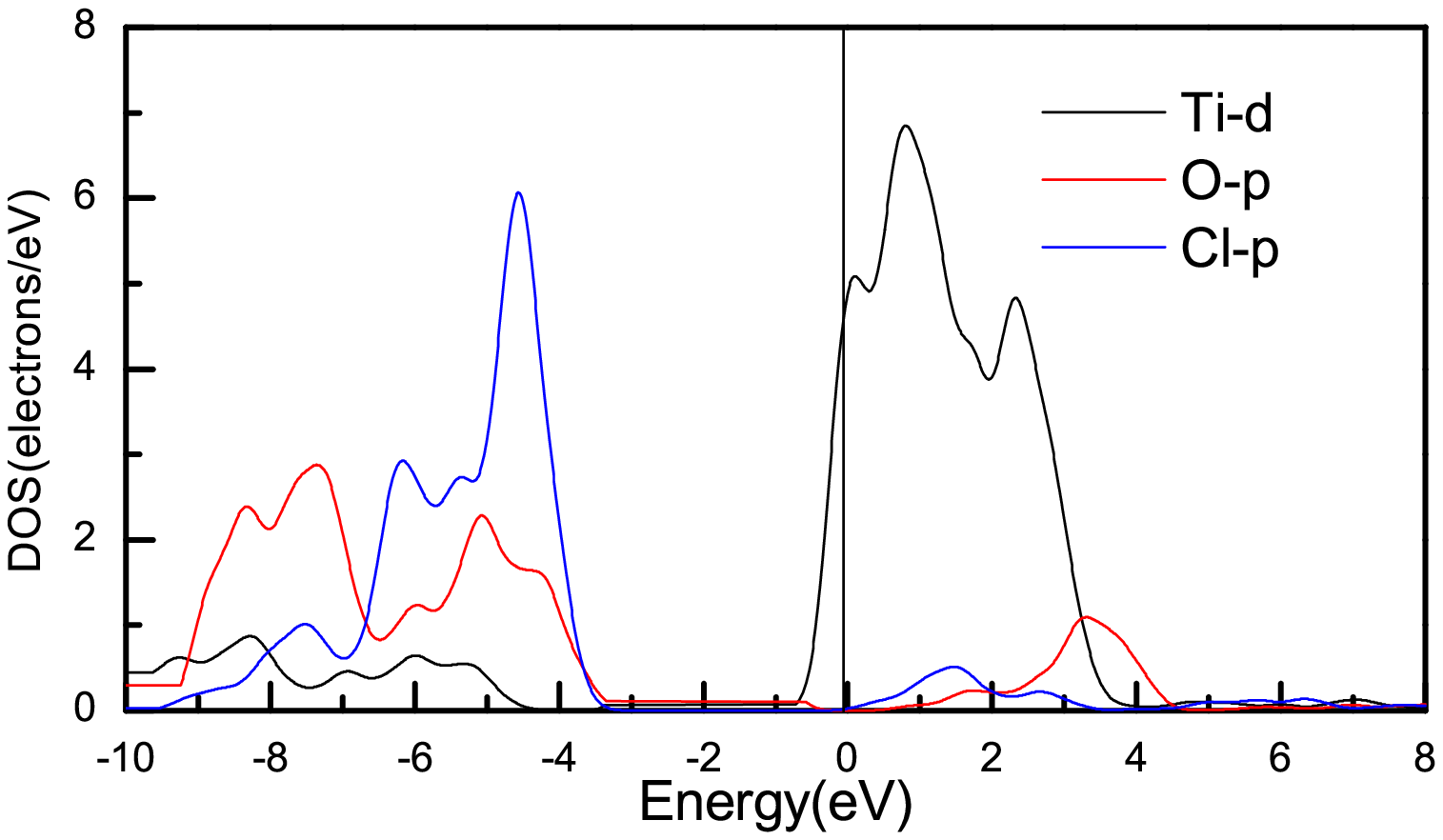}}
   \subfigure[]{\includegraphics[width=0.4\linewidth]{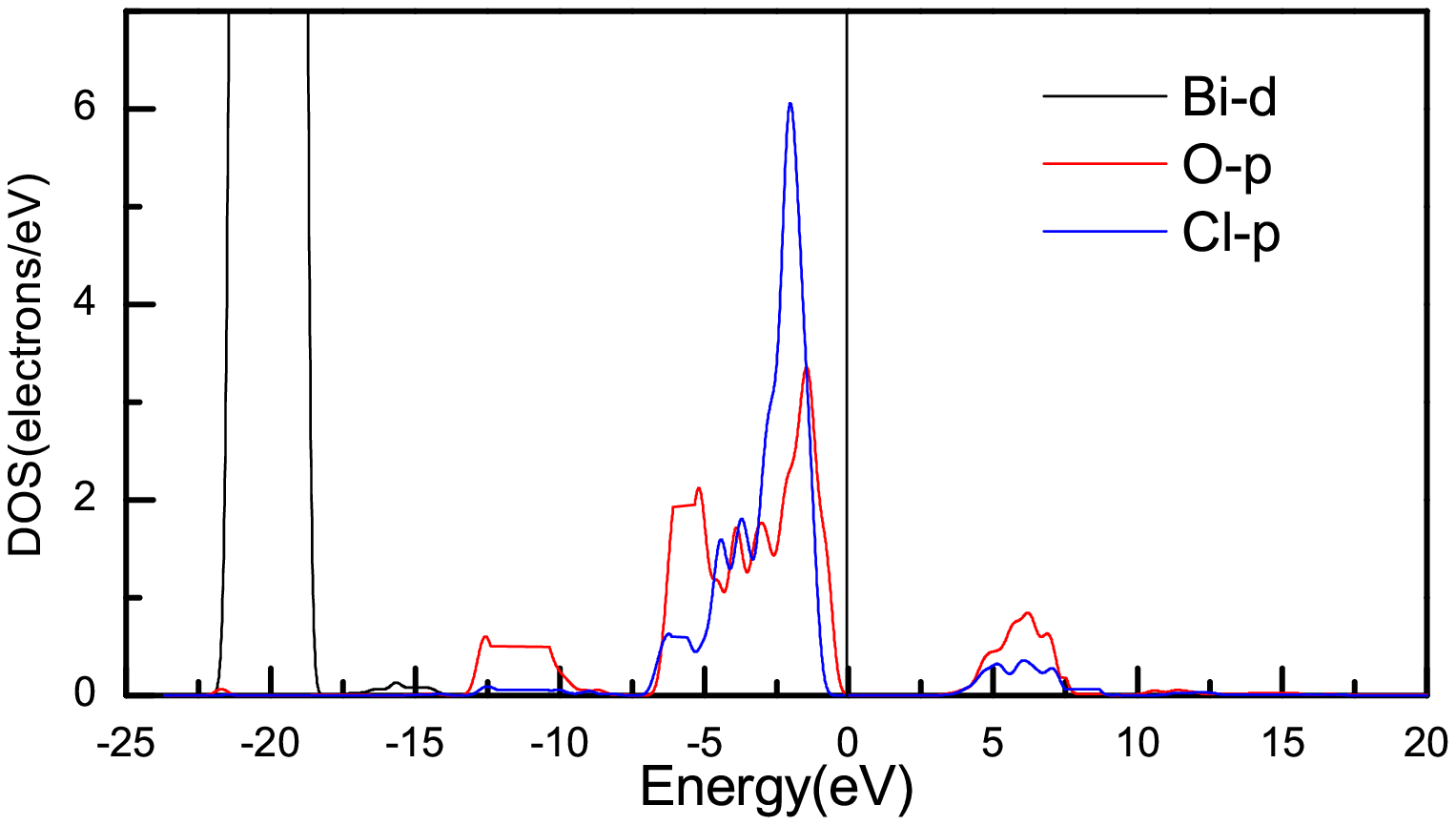}} 
   \subfigure[]{\includegraphics[width=0.4\linewidth]{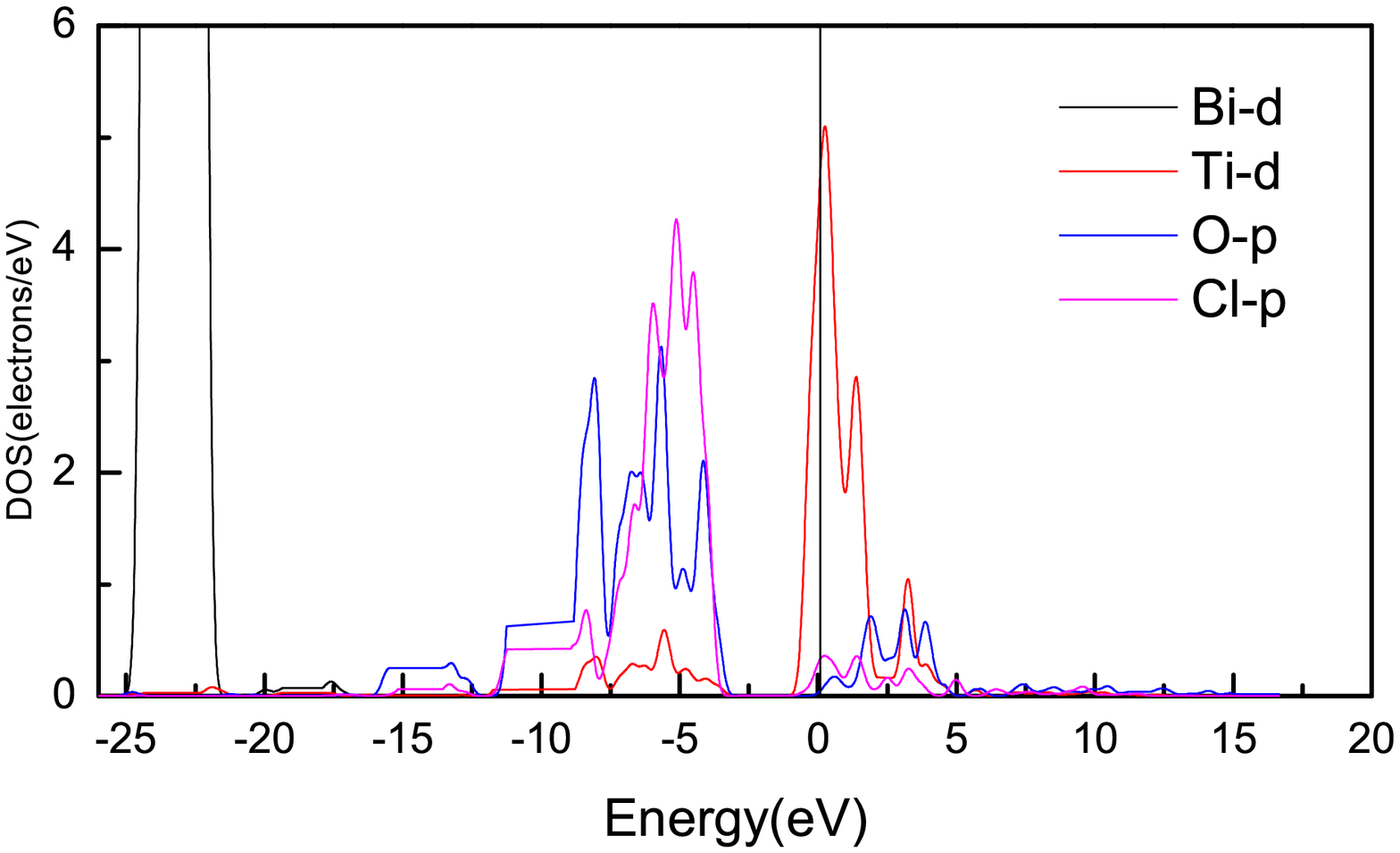}}  
   \caption{The LDA-PDOS (PDOS, partial density of states) patterns of TiOCl and BiOCl. (a) and (b) show the DOS in each level of each element as well as the total case of TiOCl and of BiOCl, respectively.
(c) and (d) show the mainly peak of level (i.e., the p-, d-, and d-level) of TiOCl and BiOCl respectively. (d) is the DOS pattern of doped crystals $Bi_{1-x}Ti_{x}OCl$ 
with doping ratio $Bi:Ti=1:1$.}
\end{figure}

\begin{figure}[!ht]
   \centering
   \subfigure[]{\includegraphics[width=0.4\linewidth]{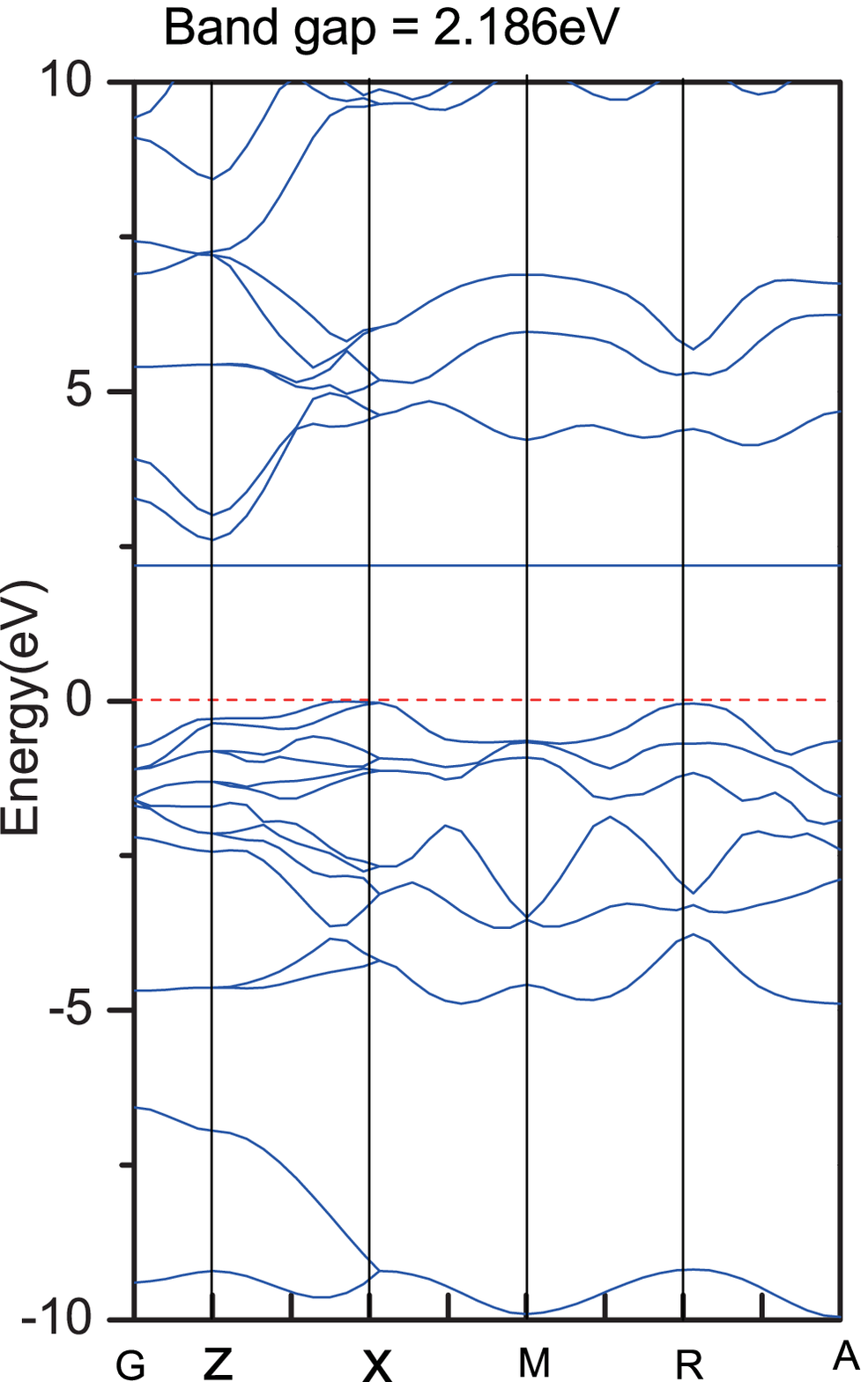}}
   \subfigure[]{\includegraphics[width=0.4\linewidth]{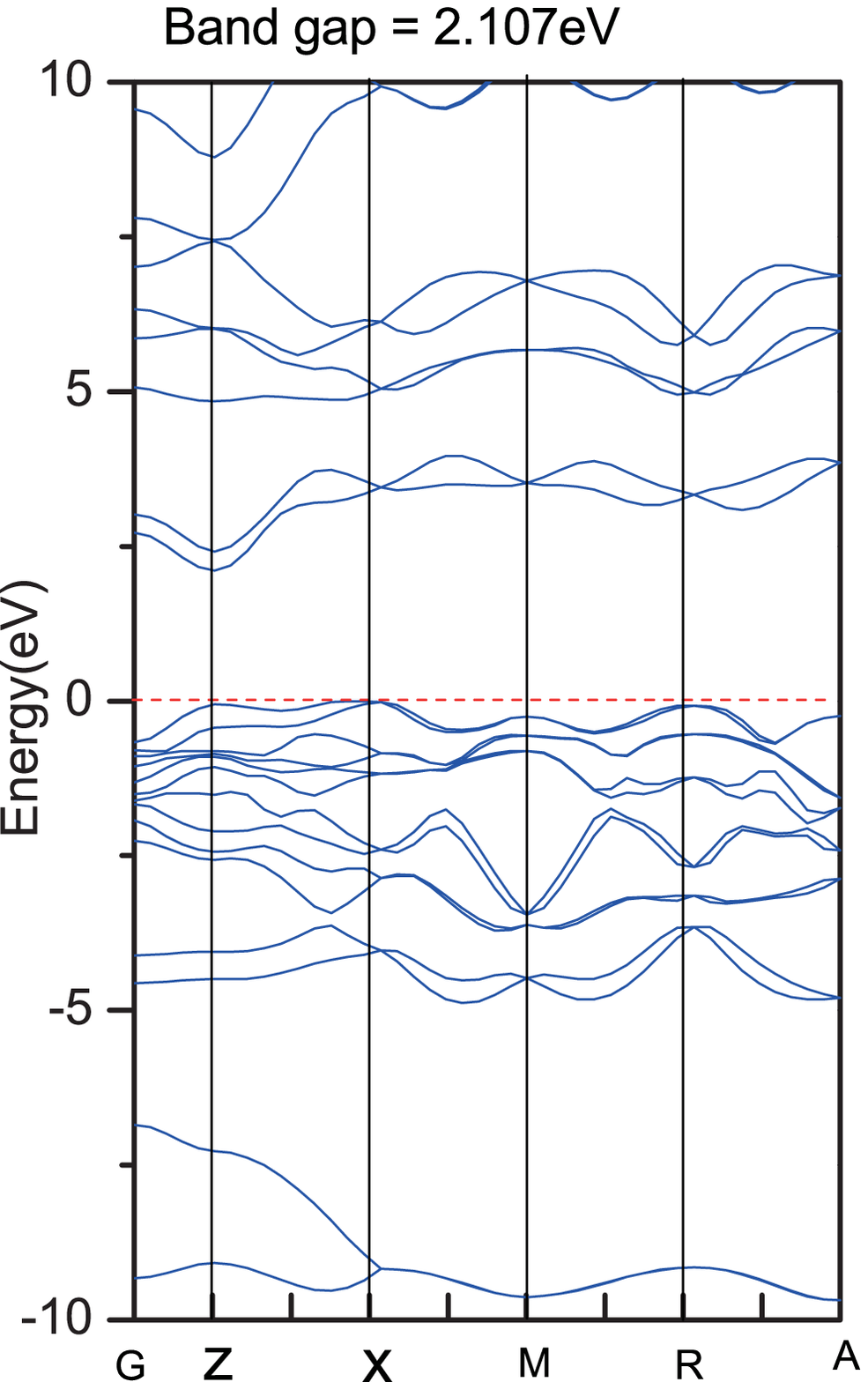}}
   \subfigure[]{\includegraphics[width=0.4\linewidth]{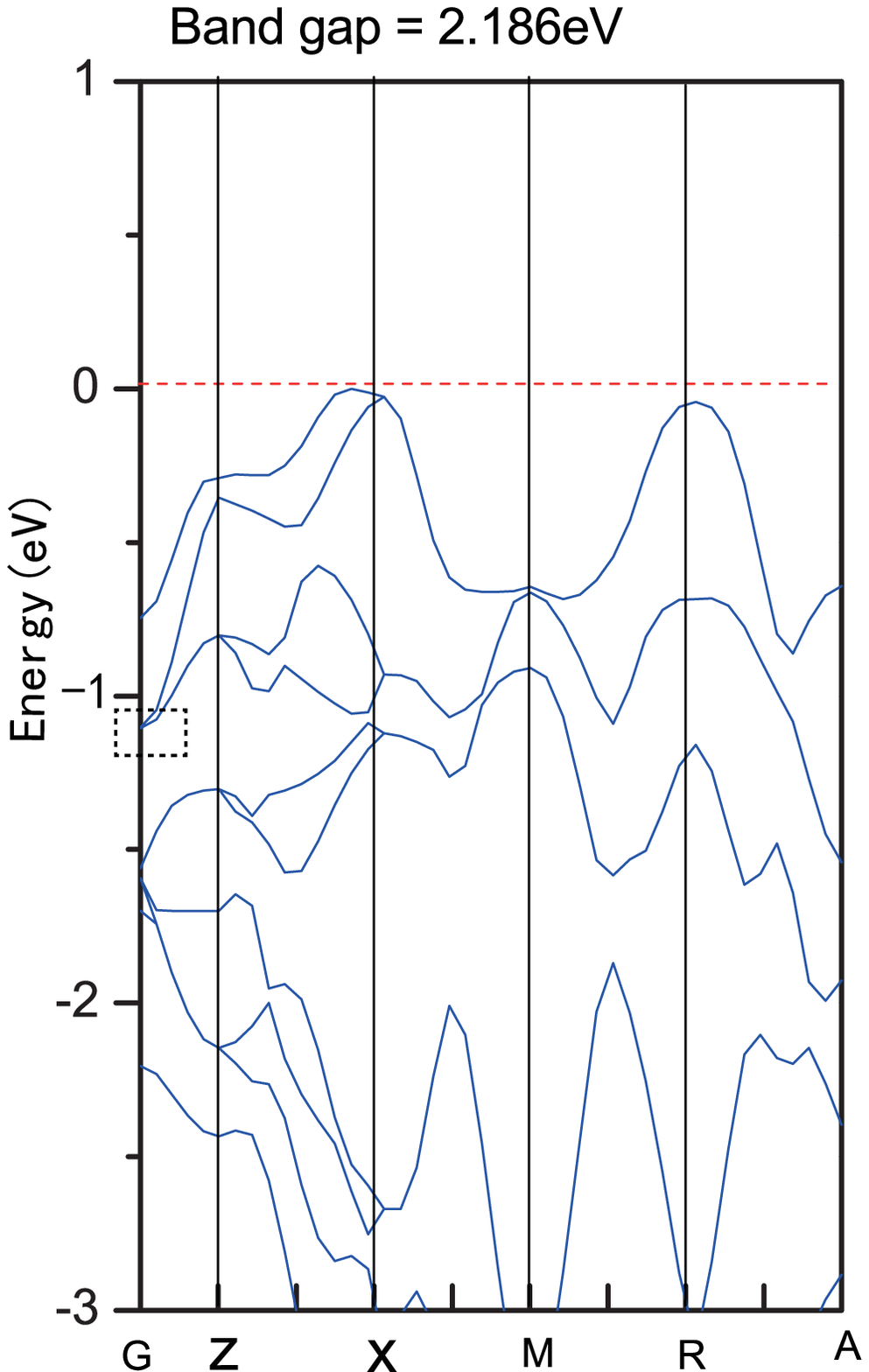}}
   \subfigure[]{\includegraphics[width=0.4\linewidth]{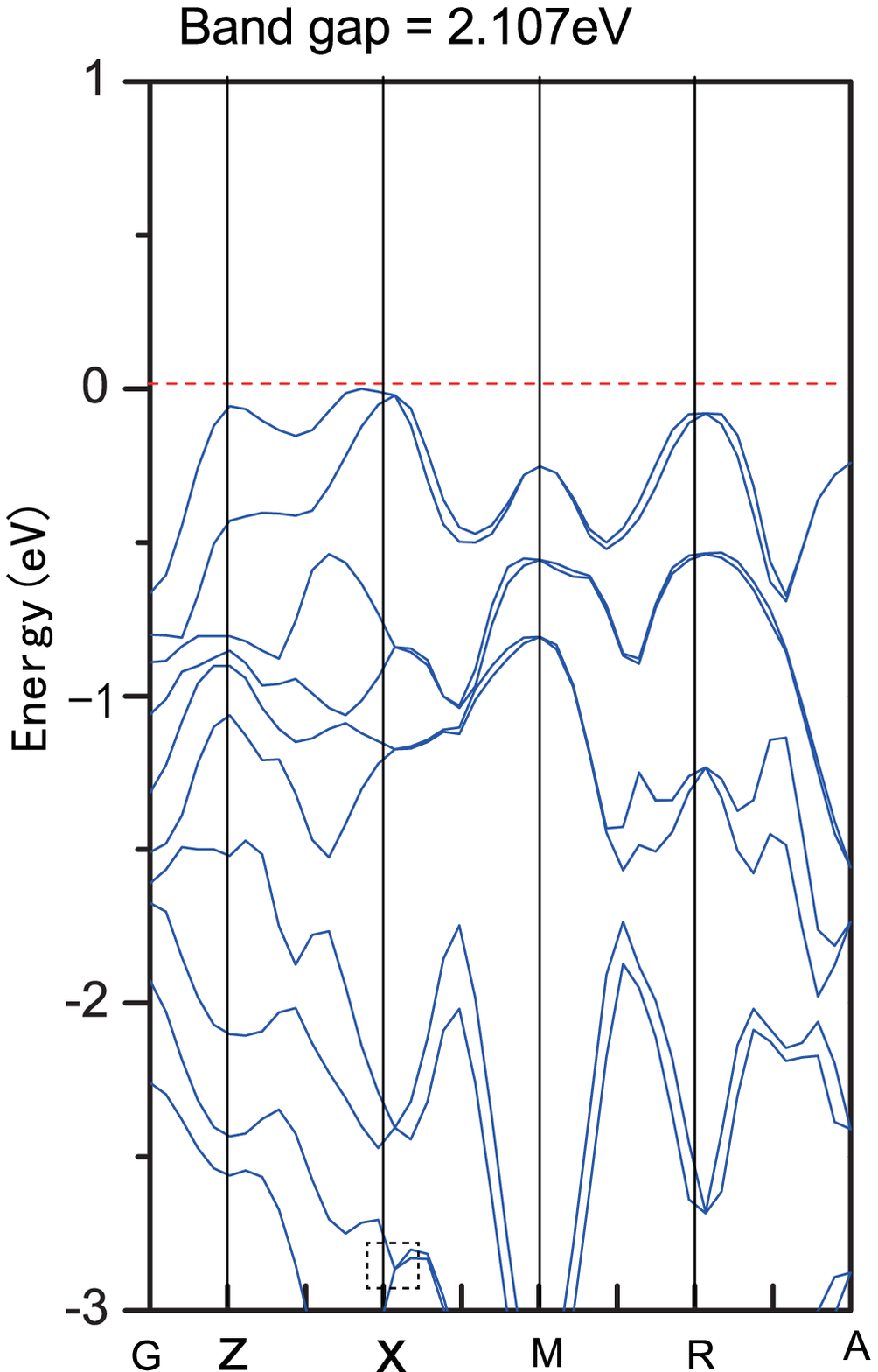}} 
   \caption{Bulk band struture for the high-symmetry phase of BiOCl. Here set the Fermi energy $E_{F}=0$. The coordinates of each point on the horizontal axis are G(0,0,0),Z(0,0,0.5),X(0,0.5,0),M(0.5,0,0),R(0,0.5,0.5),A
(0.5,0.5,0.5). (a) and (b) are the band structure of BiOCl without SOC and with SOC, respectively. (c) and (d) are the enlarged views of band structure near the fermi
energy, the dashed boxs marked in the graphs are the found three degenerate point.}
\end{figure}

\begin{figure}[!ht]
   \centering
   \subfigure[]{\includegraphics[width=0.9\linewidth]{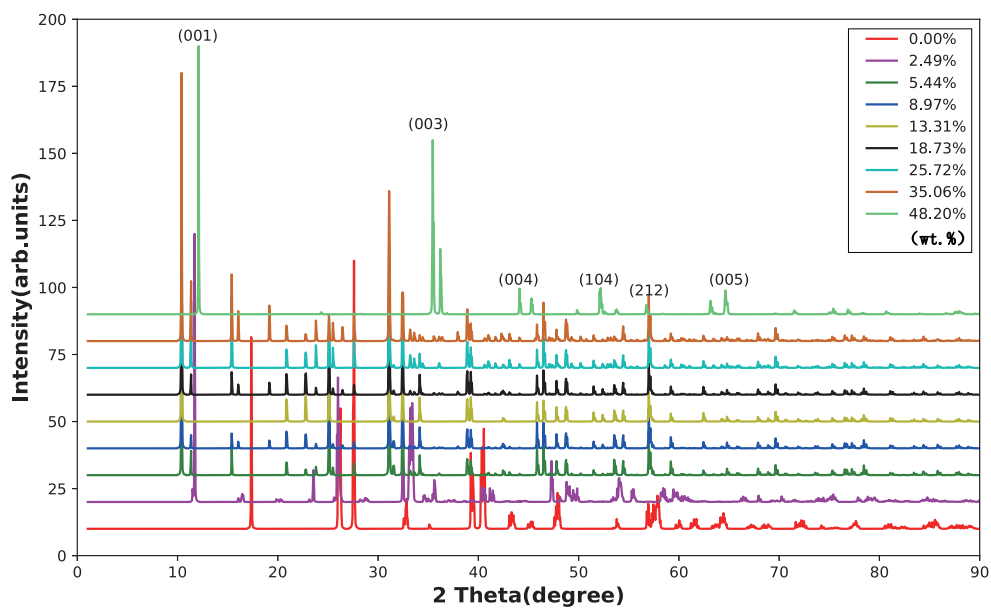}}
   \subfigure[]{\includegraphics[width=0.3\linewidth]{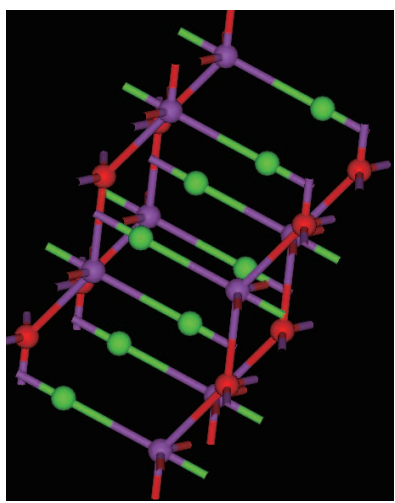}}
   \subfigure[]{\includegraphics[width=0.4\linewidth]{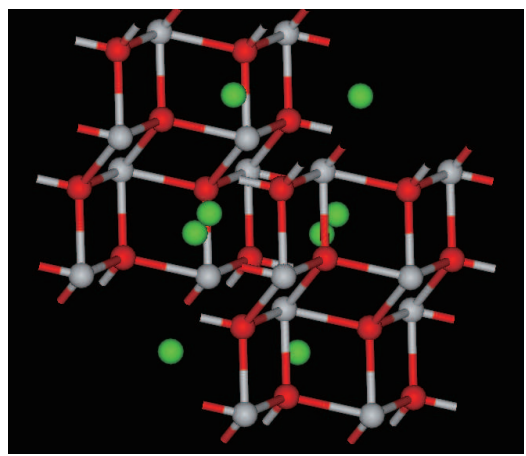}}
   \caption{(a) XRD patterns of $Bi_{1-x}Ti_{x}OCl$ with different doped weight percentage (wt.$\%$) in room temperature. 
    Note that as the doping rato increases, each line was rises by ten units more than the previous one. 
    This is for better show the changing rules of the lines, and to prevent the too much overlap of lines.
   (b) and (c) is the schematic representation of undoped BiOCl and the fully-doped TiOCl (48.20 wt.$\%$) respectively and these doping model are generated from refinement.}
\end{figure} 

\end{large}
\renewcommand\refname{References}

\end{document}